# Cepstral Coefficients for Earthquake Damage Assessment of Bridges Leveraging Deep Learning

Seyedomid Sajedi, Ph.D. Candidate,[1] Xiao Liang, Ph.D., M.ASCE[2]

[1]206 Ketter Hall, Department of Civil, Structural and Environmental Engineering, University at Buffalo SUNY, 14260-4300, Buffalo; e-mail: ssajedi@buffalo.edu
[2]242 Ketter Hall, Department of Civil, Structural and Environmental Engineering, University at Buffalo SUNY, 14260-4300, Buffalo; e-mail: liangx@buffalo.edu

**ABSTRACT**

Bridges are indispensable elements in resilient communities as essential parts of the lifeline transportation systems. Knowledge about the functionality of bridge structures is crucial, especially after a major earthquake event. In this study, we propose signal processing approaches for automated AI-equipped damage detection of bridges. Mel-scaled filter banks and cepstral coefficients are utilized for training a deep learning architecture equipped with Gated Recurrent Unit (GRU) layers that consider the temporal variations in a signal. The proposed framework has been validated on an RC bridge structure in California. The bridge is subjected to 180 bi-directional ground motion records with sampled scale factors and six different intercept angles. Compared with the benchmark cumulative intensity features, the Mel filter banks resulted in 15.5% accuracy in predicting critical drift ratios. The developed strategy for spatio-temporal analysis of signals enhances the robustness of damage diagnosis frameworks that utilize deep learning for monitoring lifeline structures.

**INTRODUCTION**

Continuous operation of bridges contributes to minimizing the impacts of natural disasters, given their key role in lifeline transportation systems. Access to reliable and timely information on the structural integrity of bridges is a critical step for effective emergency management and minimum downtime in densely populated areas. Human-based inspections have been the primary means of Structural Health Monitoring (SHM) in form of reconnaissance teams comprised of experts. Such a process can be time and resource-consuming since structural engineers are often required to visit the same structure a few times to complete an inspection. Moreover, the outcome of an inspection on the same structure may not be consistent if done by different engineering teams, making the process subjective. Due to the limitations of human-based inspections, obtaining reliable data rapidly after disasters such as earthquakes can be challenging, which further highlights the importance of automation in SHM.

The recent advances of AI have had an enormous effect on automated engineering systems' capabilities, with SHM being no exception. Many studies have been focused on visual inspections and image processing techniques that benefit from deep/machine learning algorithms (Dong and Catbas 2020; Liang 2019; Sajedi and Liang 2021). Images are valuable sources of information but face certain limits when it comes to near real-time damage diagnosis. For example, visual data might not be readily available after an earthquake, and it might only represent surface damage in concrete bridges (Kashif Ur Rehman et al. 2016). Another limitation is that quantifying structural damage is challenging or, in cases impossible, using pixel data.

Vibration-based SHM is an alternative damage diagnosis of civil infrastructure that has the potential to address these problems. Some studies have focused on using data-driven models solely relying on ground motion signals in seismic assessments (Tamhidi et al. 2020; Xu et al. 2020). However, Bridges can be instrumented with an array of accelerometers where the vibration data is uploaded into remote servers in near real-time. Instrumented structures have the potential to provide more accurate and robust information about structural damage. Vibration patterns can be utilized to obtain quantifiable measures of structural damage as they reflect dynamic structural characteristics.

The mappings between acceleration time-series and structural damage can be complicated. Data-driven SHM models have shown promising performance to find such robust mappings using the state-of-the-art deep/machine learning. Some studies have used the raw acceleration records as the input (Abdeljaber et al. 2018). Others have proposed methods to preprocess time-series before feeding them to a data-driven model (Azimi and Pekcan 2020; Mangalathu and Jeon 2020; Sajedi and Liang 2020). There have been limited studies in the literature that consider the temporal variations of input acceleration data as a part of the learning process.

Automatic speech recognition (ASR) is one area where deep learning has shown impressive success (Chan et al. 2016). Inspired by this progress, this paper investigates the potential application of Mel Filter Banks (MFBs) and Mel Frequency Cepstral Coefficients (MFCCs) for vibration-based damage diagnosis. The remainder of the paper first discusses the steps required to obtain these features. Later, a reinforced concrete bridge in California is considered to validate the potential use of the proposed signal preprocessing approach for AI-equipped vibration-based SHM.

**FEATURE EXTRACTION WITH MEL FILTER BANKS**

Acceleration records due to seismic events are commonly recorded in thousands of time steps. Using the raw vibration records as the direct input to a machine/deep learning algorithm is often challenging concerning computational costs. The Recurrent Neural Networks (RNNs) and their more advanced successors, including the Long Short-Term Memory (LSTM) and Gated Recurrent Unit (GRU), are the proper models to learn from the temporal signal variations. Given the raw vibration input size, these models could be ineffective in processing relatively long sequences. As a general rule of thumb, increasing the input tensor's complexity and size often dictates an increase in the number of learnable parameters in the data-driven model. For example, a deeper neural network architecture might be necessary. Even with unlimited computational capacities, this increase is likely to promote overfitting because the datasets for structural damage might be relatively small to calibrate millions of network weights.

An alternative approach is to preprocess the acceleration records into a new representation that A) is more compressed, B) maintains the time-variant nature of a sequence, and C) contains target sensitive features. In the following, we will explain that utilizing Mel filter banks, a representation of such characteristics can be obtained for vibration-based damage diagnosis. This process is initially described for a single record. We will later explain how to consider the information from several sensors with multiple channels.

The first step is to break the acceleration signal into a series of overlapping time frames (Figure 1.a and 1b). Frame's length and stride are the two hyperparameters selected based on the signal characteristics such as Sampling Rate (*SR*). These values are often in the scale of milliseconds for speech signals given substantially higher *SR*. We noticed that tensors representing human voices are often much larger than the existing earthquake records. For example, an audio signal typically has sampling rates in the order 8-16 kHz. However, ground motions in the PEER NGA West 2 database (Ancheta et al. 2014) are often recorded with approximately 50-500 Hz sampling rates (i.e., times steps varying between 0.002-0.02s). Based on such observations, a window length of 1 s with the stride of 0.4 s is selected to discretize the signal. The number of frames or windows ($N_w$) will depend on the ground motion duration and is variable for each earthquake realization.

Each time frame includes the acceleration amplitudes in the time domain. A better representation can be obtained by creating the periodogram of the frames (Figure 1c). This is possible by taking the Fast Fourier Transform (FFT) and extracting $N_{FFT}$ coefficients for each frame. $N_{FFT}$ =512 is considered, and a signal will be padded with zeros if the number of data points in a frame is less than $N_{FFT}$ depending on an earthquake record's *SR*. The periodogram of a frame is obtained by considering the magnitude of these complex coefficients ($|FFT(a_i)|$). The following equation is used to obtain points that construct this spectrum ($X_f$):

$$X_f = \frac{|FFT(a_i)|^2}{N_{FFT}}. \quad \text{(Eq. 1)}$$

Similar to the accelerations series, the periodogram includes a substantial amount of $X_f$'s for each frame in the frequency domain. The cepstral coefficients are a compressed representation of the periodogram where instead of individual values, different intervals of frequencies are considered. To this end, the Mel scale is proposed to define such intervals (Davis and Mermelstein 1980). Each Hertz point ($f_i$) can be converted into an equivalent Mel value ($M_i$), using the following equation:

$$M_i = 2595\log(1+\frac{f_i}{700}). \quad (Eq.\ 2)$$

Based on the Nyquist theorem, the Fourier coefficients expand from 0 to *SR*/2 Hz. As a result, the periodogram's horizontal axis is bounded between these two values, which will be evaluated for $N_{fl}$ intervals. Cepstral coefficients are indicators of the amount of energy in different frequency ranges. These indicators are measured by independently multiplying $N_{fl}$ triangular filters by the periodogram (Figure 1d). An MFB coefficient is calculated by taking the logarithm of this multiplication for a given frequency range (Figure 1.e). The filter frequencies are obtained by considering $N_{fl}$+2 equally-spaced $M_i$ points between 0 and the corresponding Mel point for *SR*/2. The peak magnitude of each filter is set to one. The sensor data is digital, and the filter points may not precisely match the periodogram's Hertz points. This issue is addressed with further technical details by Lyons (2020).

After determining the filters, $N_{fl}$ coefficients can be obtained from a periodogram. Repeating this process for all frames in a signal results in a 2-dimensional tensor representation with the size of $N_w \times N_{fl}$. It is common in ASR to perform an additional step by taking the Discrete Cosine Transform (DCT) of the MFB for each frame because the coefficients could be correlated in each frame. Taking the DCT will yield MFCC coefficients, which is also a tensor of the same $N_w \times N_{fl}$ shape (Figure 1.f). In the following sections, we will compare MFB and MFCC for the proposed case study.

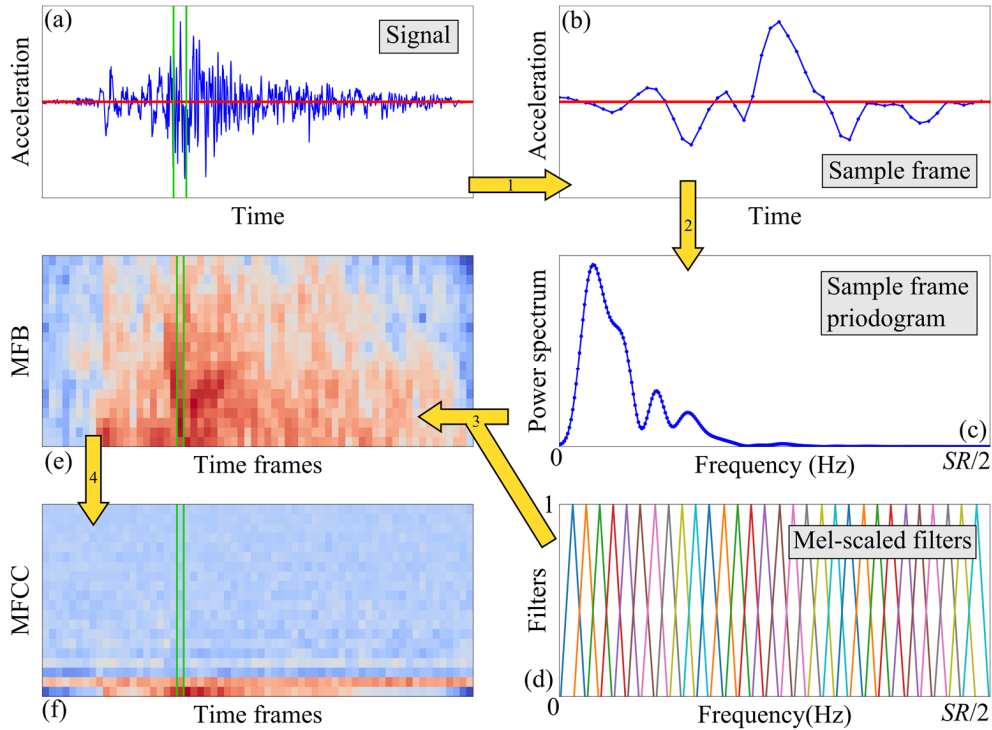

**Figure 1. Different steps to extract MFB and MFCC from an acceleration time-series**

## VALIDATION CASE STUDY

The signal processing method will be evaluated for the damage diagnosis of the bridge structure (on-ramp overcrossing) in Jack Tone road, located in Ripon, California. The reinforced concrete bridge specification and model are adopted from the PEER report by Kaviani et al. (2014). The circular pier is modeled with nonlinear fiber sections, while the bridge deck is assumed to be linear elastic. The finite element model also considers several spring elements to model the shear key, backfill, and the expansion joint, as shown in Figure 2.

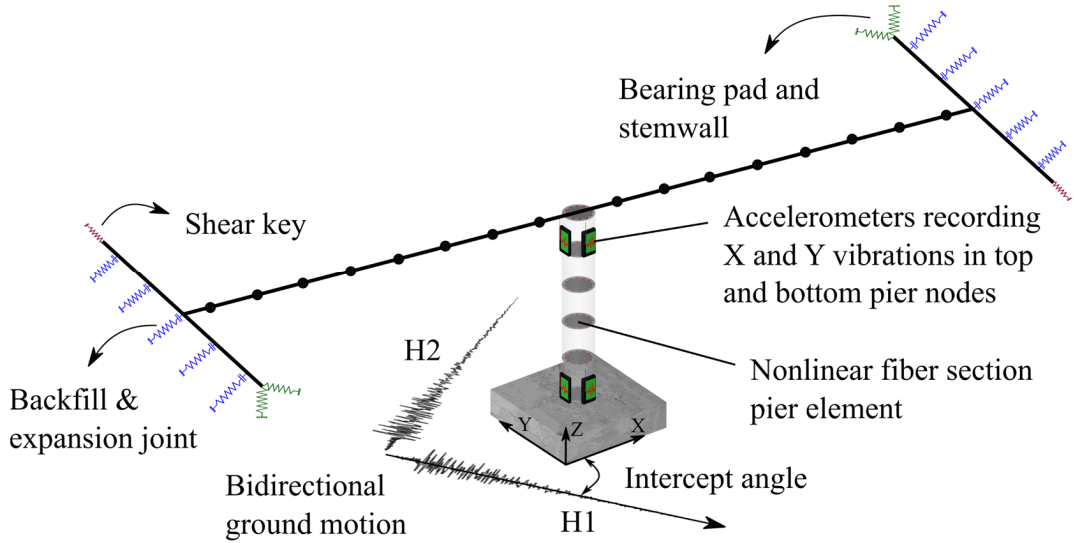

**Figure 2. Bridge model specifications**

180 earthquake Ground Motion (GM) records are selected by considering an M7 earthquake scenario and 30 PGV intensities (Liang and Mosalam 2020; Sajedi and Liang 2020). Seismic hazard probabilities for each intensity are adopted from the CB-2014 attenuation model (Campbell and Bozorgnia 2010). It is also assumed that the 3D structure is subjected to bi-directional ground motions with 0, 60, 90, 120, 150 degrees of intercept angle with respect to the bridge deck (Liang et al. 2016). These assumptions yield 32,400 nonlinear response history analyses that are performed in OpenSees. The simulations are conducted on 1,000 computation nodes at the University at Buffalo's Center for Computational Research (CCR).

The dataset is split into training, validation, and testing sets based on the ground motion criteria. 80% of the GMs (144), including all intercept angles and intensities, are randomly selected for training. The remaining 20% is used for testing. 10% of the training data is also used as the validation set for tuning the hyperparameters of the deep learning models defined in the next section.

It is assumed that the bridge pier is instrumented with two accelerometers located at the top and bottom that record accelerations in global X and Y directions. The processed vibration input is, therefore, comprised of four $N_w \times N_{fl}$ tensors. The majority of MFCC coefficients for higher frequency ranges are close to zero and might be ignored in the damage diagnosis process.

Therefore, the first 8 coefficients of MFB and MFCC are considered for each sensor channel. Furthermore, the processed masks are combined into a single tensor in a two-step process. This procedure is shown for sample MFB features in Figure 3. First, the bottom mask is subtracted from the one at the top in X and Y directions. Next, the difference tensor in the Y direction is vertically stacked to the one in the X direction, forming a single tensor of shape $N_w \times 16$.

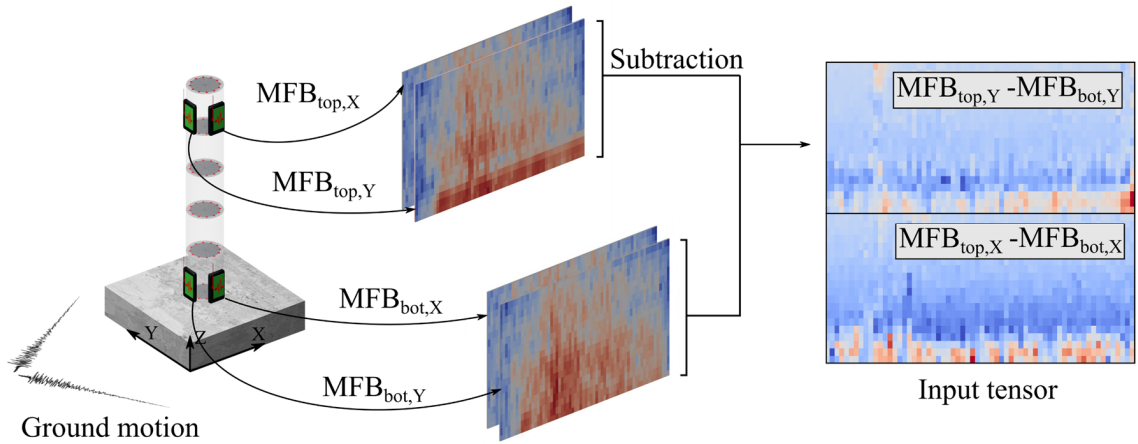

Figure 3. Constructing a single MFB input tensor from multiple sensors

A benchmark input feature is also considered further to evaluate the quality of the proposed signal processing technique. Cumulative intensity measures ($I^\eta$) are another type of features utilized in data-driven SHM and shown promising performance (Liang et al. 2018; Sajedi and Liang 2020; Eltouny and Liang 2021). This feature can be expressed as:

$$I^\eta = \int_0^{t_g} |a(t)|^\eta dt .  \qquad (\text{Eq. 3})$$

where $t_g$ is the duration of the ground motion, $a(t)$ is the acceleration magnitude at time $t$, and $\eta$ is a tuning hyperparameter. To enrich the benchmark feature space, $I^\eta$ is measured for all sensor channels while considering multiple values of $\eta \in \{0.2, 0.4, 0.6, ..., 1.8, 2.0\}$. The extracted $I^\eta$ values are stacked into a single vector as the benchmark input data. The issue with $I^\eta$ is that temporal information in a signal is lost after the integration. In the following sections, MFB, MFCC and $I^\eta$ will be compared regarding their capability for damage prediction.

**DEEP LEARNING ARCHITECTURE**

The RNN architectures are an optimal choice to deal with sequence data. There are two essential gains from using an RNN-based model. First, the learning process will consider the temporal variations in the input due to feedback connections. Moreover, the RNN-based models have advanced to overcome computational challenges after several years of research (Hochreiter and Schmidhuber 1997). The second advantage is that the data-driven model can be trained end-to-end for vibration response of earthquakes with different durations.

The deep learning model in this paper utilizes ReLU-activated GRUs (Chung et al. 2014), which showed a better performance compared with a similar architecture using LSTMs. The neural network is comprised of the temporal and bottleneck parts. The temporal part begins with a masking layer. Input MFB or MFCC masks are initially padded with zero such that all events have a constant number of 500 frames. The masking layer will automatically ignore the padded frames in the following GRU feedback connections. Several GRU layers with an increasing number of extracted features come after the masking layer. This will conclude the temporal part of the architecture.

The bottleneck includes a series of fully connected (dense) layers with tanh activation function. We observed that compared with ReLU, tanh is substantially more stable and yields better results. A dropout ratio of 5% is used for the first dense layer to alleviate overfitting. In this study, it is intended to estimate the drift values as a regression problem. Therefore, the architecture is concluded with a single-output dense layer and sigmoid activation. Further details can be found in Table 1. The target output values are amplified by 10 for numerical stability because the drift ratios do not exceed 10% in the whole dataset.

**Table 1. The Deep learning architecture for drift regression**

| Layer | Activation |
|---|---|
| The temporal part of the architecture ||
| Masking | - |
| GRU(50) | ReLU |
| GRU(60) | ReLU |
| GRU(70) | ReLU |
| GRU(80) | ReLU |
| GRU(90) | ReLU |
| GRU(100) | ReLU |
| The bottleneck part of the architecture ||
| Dense(2000) | tanh |
| Dropout(5%) | - |
| Dense(2000) | tanh |
| Dense(1) | sigmoid |

The presented architecture in Table 1 is used for MFB and MFCC inputs. Since $I^\eta$ features are already processed by the integration shown in Eq. 3, the deep learning architecture's temporal part is not required to process these features in drift regression. Therefore, the benchmark input will be directly fed to the bottleneck part of the architecture. It should be noted that the majority of the networks learnable parameters are located in the bottleneck. The temporal part includes less than 10% of the total number of these parameters (4,403,251).

Keras API (Chollet 2015) is used for this implementation. Mean Absolute Error (MAE) is considered the loss function, minimized using the Nadam optimizer. Three models are trained individually with a batch size of 1200 realizations. The benchmark model's learning rate is reduced from the default value of 2.0e-3 to 1.0e-6 for better and more stable convergence. Both models are trained until the validation loss stopped improving with the patience of 10 epochs.

**RESULTS AND DISCUSSION**

Testing MAE metric for the three types of features is provided in Table 2. It can be seen that the MFB is 15.5% more accurate compared with the benchmark $I^\eta$. It is also observed that further processing the MFBs by taking the DCT does not improve the performance. Therefore, MFBs are better choices for damage diagnosis than MFCCs in this case-study and unlike most ASR applications. Figures 1.e and 1.f are examples of MFB and MFCC tensors, which imply why this observation makes sense. After taking the DCT, most MFCC coefficients are close to zero. In contrast, MFB provides a more detailed representation of variations in the signal.

**Table 2. Testing performance metrics**

| Feature type | MAE (%) |
|---|---|
| MFB | 0.5129 |
| MFCC | 0.6771 |
| $I^\eta$ | 0.6067 |

The regression accuracy of the models based on MFB and $I^\eta$ features is furthered investigated in the scatter plots of Figure 4. The ground truth drift values and the corresponding absolute errors are illustrated in Figure 4.b. To track these variations in error, a polynomial fit of absolute error with respect to the drift values is provided (solid lines in the graphs). It can be observed that the proposed MFB model yields a lower average error. The improvements are more significant in higher drift ratios, where the structure has experienced higher levels of nonlinearity. Therefore, it can be concluded that the GRU-equipped model has better learning capacities compared with the benchmark.

**CONCLUSION**

Rapid condition assessment of bridges is an essential step in recovering from natural disasters such as earthquakes. The information obtained by such assessments can guarantee the safe and continuous operation of lifeline transportation systems. Inspired by the latest advances in Automatic Speech Recognition (ASR) and AI, we investigated a novel deep learning framework for automatic vibration-based damage diagnosis of bridges.

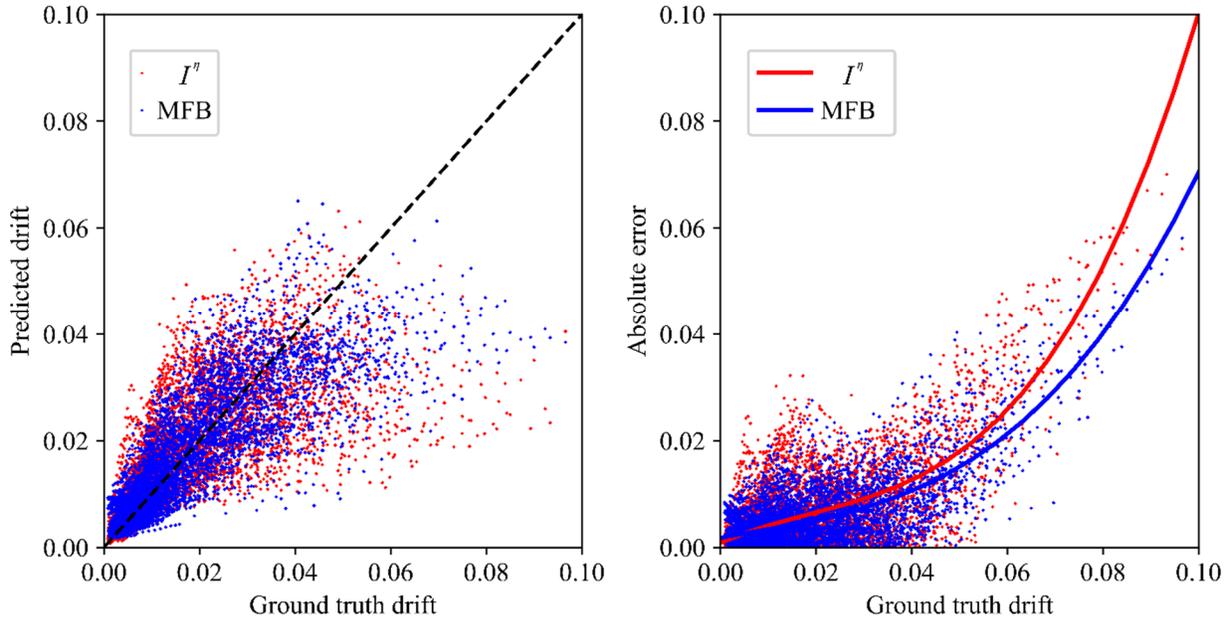

**Figure 4. Scatter plots comparing the regression performance of MFB vs $I^{\eta}$**

Two signal preprocessing strategies are proposed based on Mel-scaled filters to obtain MFB and MFCC masks. We have shown that using a deep learning architecture equipped with GRUs, the temporal variations in a signal can be considered during the training process. The methodology is evaluated on a bridge structure in California subjected to 32,400 earthquake response simulations. The proposed approach showed 15.5% more accuracy compared with the benchmark $I^{\eta}$ features for damage diagnosis of bridges. MFB features have shown a promising performance and robustness for automated SHM of bridge infrastructure. Future studies will focus on an enhanced Mel scale and filter banks specifically tailored for earthquake damage detection rather than ASR.

**REFERENCES**

Abdeljaber, O., Avci, O., Kiranyaz, M. S., Boashash, B., Sodano, H., and Inman, D. J. (2018). "1-D CNNs for structural damage detection: verification on a structural health monitoring benchmark data." *Neurocomputing*, 275, 1308-1317.

Ancheta, T. D., Darragh, R. B., Stewart, J. P., Seyhan, E., Silva, W. J., Chiou, B. S.-J., Wooddell, K. E., Graves, R. W., Kottke, A. R., and Boore, D. M. J. E. S. (2014). "NGA-West2 database." 30(3), 989-1005.

Azimi, M. and Pekcan, G. (2020). "Structural health monitoring using extremely compressed data through deep learning." *Computer‐Aided Civil Infrastructure Engineering*, 35(6), 597-614.

Campbell, K. W. and Bozorgnia, Y. (2010). "A ground motion prediction equation for the horizontal component of cumulative absolute velocity (CAV) based on the PEER-NGA strong motion database." *Earthquake Spectra*, 26(3), 635-650.

Chan, W., Jaitly, N., Le, Q., and Vinyals, O. "Listen, attend and spell: A neural network for large vocabulary conversational speech recognition." *Proc., 2016 IEEE International Conference on Acoustics, Speech and Signal Processing (ICASSP)*, IEEE, 4960-4964.

Chollet, F. (2015). "Keras."

Chung, J., Gulcehre, C., Cho, K., and Bengio, Y. (2014). "Empirical evaluation of gated recurrent neural networks on sequence modeling." *arXiv preprint*.

Davis, S. and Mermelstein, P. (1980). "Comparison of parametric representations for monosyllabic word recognition in continuously spoken sentences." *IEEE transactions on acoustics, speech, signal processing*, 28(4), 357-366.

Dong, C.-Z. and Catbas, N. (2020). "A review of computer vision-based structural health monitoring at local and global levels." *Structural Health Monitoring*.

Eltouny, K. A. and Liang, X. (2021). "Bayesian-Optimized Unsupervised Learning Approach for Structural Damage Detection." *Computer-Aided Civil & Infrastructure Engineering*.

Hochreiter, S., and Schmidhuber, J. J. N. c. (1997). "Long short-term memory." 9(8), 1735-1780.

Kashif Ur Rehman, S., Ibrahim, Z., Memon, S. A., and Jameel, M. (2016). "Nondestructive test methods for concrete bridges: A review." *Construction and Building Materials*, 107, 58-86.

Kaviani, P., Zareian, F., and Taciroglu, E. (2014). "Performance-Based Seismic Assessment of Skewed Bridges, Report PEER 2014/01." Pacific Earthquake Engineering Research Center, University of California Irvine.

Liang, X. (2019). "Image-Based Post-Disaster Inspection of Reinforced Concrete Bridge Systems Using Deep Learning with Bayesian Optimization." *Computer-Aided Civil & Infrastructure Engineering*, 34(5), 415-430.

Liang, X. and Mosalam, K.M. (2020). "Ground Motion Selection and Modification Evaluation on Highway Bridges Subjected to Bi-directional Horizontal Excitation." *Soil Dynamics & Earthquake Engineering*, 130, 105994.

Liang, X., Mosalam, K.M., and Günay, S. (2016). "Direct Integration Algorithms for Efficient Nonlinear Seismic Response of Reinforced Concrete Highway Bridges." *Journal of Bridge Engineering*, 21(7), 04016041.

Liang, X., Mosalam, K., and Muin, S. (2018). Simulation-based data-driven damage detection for highway bridge systems. *Proc. 11th Nat. Conf. Earthquake Eng. (NCEE)*.

Lyons, J. (2020). "Mel Frequency Cepstral Coefficient (MFCC) tutorial [Practical Cryptography]." <URL: http://practicalcryptography. com/miscellaneous/machine-learning/guide-mel-frequency-cepstral-coefficients-mfccs>.

Mangalathu, S. and Jeon, J.-S. (2020). "Ground Motion-Dependent Rapid Damage Assessment of Structures Based on Wavelet Transform and Image Analysis Techniques." 146(11), 04020230.

Sajedi, S. O. and Liang, X. (2020). "A data-driven framework for near real-time and robust damage diagnosis of building structures." *Structural Control & Health Monitoring*, 27(3), e2488.

Sajedi, S. O. and Liang, X. (2020). "Vibration-based semantic damage segmentation for large-scale structural health monitoring." *Computer-Aided Civil Infrastructure Engineering*, 35(6), 579-596.

Sajedi, S. O. and Liang, X. (2020). "Dual Bayesian Inference for Risk-Informed Vibration-Based Damage Diagnosis." *Computer-Aided Civil & Infrastructure Engineering*.

Sajedi, S. O. and Liang, X. (2021). "Uncertainty-Assisted Deep Vision Structural Health Monitoring." *Computer-Aided Civil & Infrastructure Engineering*, 36(2), 126-142.

Tamhidi, A., Kuehn, N., Ghahari, S. F., Taciroglu, E., and Bozorgnia, Y. (2020). "Conditioned Simulation of Ground Motion Time Series using Gaussian Process Regression."

Xu, Y., Lu, X., Cetiner, B., and Taciroglu, E. (2020). "Real-time regional seismic damage assessment framework based on long short‐term memory neural network." *Computer-Aided Civil Infrastructure Engineering*.